\documentclass[11pt]{article}
\usepackage{moriond,epsfig,axodraw}
\usepackage{amsmath}
\usepackage{amssymb}

\def\Journal#1#2#3#4{{#1} {\bf #2}, #3 (#4)}

\def\PLB{{\em Phys. Lett.}  B}

\def\PRD{{\em Phys. Rev.} D}

\def\mco{\multicolumn}

\def\ra{\rightarrow}

\def\ko{K^0}

\def\be{\begin{equation}}
\def\ee{\end{equation}}
\def\bea{\begin{eqnarray}}
\def\eea{\end{eqnarray}}

\begin{document}
\vspace*{4cm}
\title{LEPTOGENESIS FROM RIGHT-HANDED NEUTRINO DECAYS
TO RIGHT-HANDED LEPTONS}

\author{ T. HAMBYE }

\address{Instituto de F\'\i sica Te\'orica,
Universidad Aut\'onoma de Madrid,\\
Cantoblanco,
Spain}

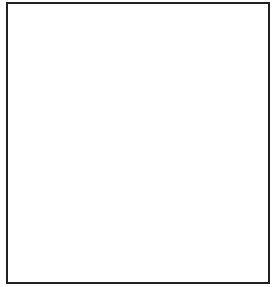
\maketitle\abstracts{We investigate what would be the consequences for 
leptogenesis of the existence of a charged $SU(2)_L$ 
singlet scalar $\delta^+$. If such a scalar particle exists it 
allows the right-handed neutrinos
to couple not only to left-handed lepton and Higgs doublets as in ordinary 
leptogenesis, but also to a right-handed charged lepton and a $\delta^+$.
This provides a new source of leptogenesis which can be successful in 
a non-resonant way at scales as low as TeV.
The incorporation of this scenario in left-right symmetric and unified models
is discussed.}

\section{Introduction}

The leptogenesis mechanism \cite{fy} provides a particularly simple and well motivated 
explanation for the origin of the baryon asymmetry of the universe. Its motivation is based 
on the recent discovery of the neutrino masses and the fact that these 
masses are most presumably associated with lepton number violation. The 
lepton number 
violation associated to the neutrino masses can create a lepton number asymmetry 
at high temperature in the universe.  
This results in the creation of a baryon asymmetry from the lepton to baryon 
conversion induced by the Standard Model sphalerons associated to 
the $B+L$ anomaly.
The most straightforward and presumably most attractive way to induce 
the neutrino masses and leptogenesis is the type-I seesaw model \cite{seesaw} in which 
the lepton asymmetry can be produced from the effect of the L violating right-handed (RH) neutrino Majorana masses in the RH neutrino decays.

In a generic way leptogenesis from decay of RH neutrinos can 
easily lead to the 
observed baryon asymmetry of the universe, i.e. to a baryon to 
entropy density of the universe equal to \cite{spergel}  $n_B/s=9 \cdot 10^{-11}$. However there are at least two criticisms one could make on this framework. The first is more pragmatic
than theoretical: due to the 
smallness of the neutrino masses, in a generic way
type-I leptogenesis can work only at a very high 
scale \cite{bcst,th,di} (i.e.~if \cite{di,bdp,gnrrs} $M_{N_1} \gtrsim 4 \cdot 10^8$
GeV where $N_1$ is the lightest RH neutrino). Beside the 
fact that this bound is 
in tension with the gravitino constraint in 
supergravity theories, basically it implies 
that leptogenesis could {\it never} be tested directly.
The second criticism is more theoretical: 
at such scale far beyond the reach of present accelerators 
we have of course no 
guarantee at all that the right-handed neutrinos
exist and that they provide the only source of lepton number violation. 
It turns out that there are quite a few other ways to induce 
successful leptogenesis 
at a high scale.
Leptogenesis is a mechanism which in this sense works even too easily.
For example, an alternative source of 
neutrino masses which can lead to successful leptogenesis~\cite{hs,ak,sod}, 
and which is well motivated in unified theories such as SO(10), Pati-Salam or 
left-right model, is the type II seesaw. It involves the interactions 
of a heavy $SU(2)_L$ triplet Higgs $\Delta_L$. Other 
seesaw possibilities of successful leptogenesis arise if 
there are two or more 
heavy Higgs triplets \cite{ms,hms} or if self-conjugate triplets of 
fermions $\Sigma$ exist \cite{hlnps}.
These models also work generically only at a high scale, 
i.e.~if \cite{hrs,hs,hms}  $M_{\Delta_L} \gtrsim 2.5 \cdot 10^{10}$ GeV or 
if \cite{hlnps} $M_{\Sigma_1} \gtrsim 1.5 \cdot 10^{10}$ GeV.

Beside looking at the various leptogenesis possibilities at a high scale, in 
the absence of any real possibilities to distinguish 
these models experimentally, another important question to investigate
is to see more phenomenologically what are the basic mechanisms and 
interactions which 
could induce successful 
leptogenesis at a directly testable low scale, even if in this case there 
is always a price to pay in terms of assumptions to be made (particle content 
extended and/or fine-tuning assumed, naturalness in grand unified theories, 
relaxation of the links between neutrino mass constraints and leptogenesis). 
At low scale too, it turns out that there are several possibilities to 
induce leptogenesis, resonant ones~\cite{fps,pil2} or non-resonant 
ones \cite{th,bhs,aal}.
In this talk I want to emphasize the fact that low scale leptogenesis doesn't
necessarily require to assume a quasi-degeneracy of the heavy particle mass spectrum
or to assume two sources of lepton number violation, one 
for neutrino masses 
and a different one for leptogenesis.  
I present  a new mechanism of leptogenesis which can work at low scale,
where a) neutrino masses are induced as in the type-I model, 
b) the source of lepton number 
violation remains the same  
(i.e.~the Majorana masses of RH neutrinos $N_i$), and c) the 
decays of the RH neutrinos are also at the origin of leptogenesis, but 
where d) the 
interactions driving dominantly the decays of the right-handed neutrinos,
instead of involving the left-handed Standard Model (SM) leptons, involve 
the right-handed SM leptons.
The price to pay with respect to high energy models is that, in 
order that the right-handed neutrinos can decay to right-handed leptons,
a new particle has to be assumed to exist, a $SU(2)_L$ singlet charged 
scalar $\delta^+$. I show that if this particle exists leptogenesis
can be implemented in a very simple way even at scales as low as $\sim$ TeV.
This work is based on a collaboration with Michele Frigerio and Ernest Ma \cite{FHM}.

\section{The Model}

The minimal implementation of our mechanism requires that, 
in addition to the SM particles, there exist two or more RH neutrinos $N_i$ 
and a charged scalar $SU(2)_L$ singlet $\delta^+$. From this particle content one can write down the most general lagrangian and the interactions involving the $\delta^+$: 
\begin{eqnarray}
{\cal L} &\owns& -M^2_\delta \delta^{+\dagger} \delta^+
+\left[-\frac{1}{2} M_{N_{i}} N^{T}_{iR} C N_{iR} 
- H^\dagger \bar{N}_{iR}  (Y_N)_{ij} \psi_{jL} \nonumber \right.\\
& &  - (Y_R)_{ij} N_{iR}^T C 
\delta^+ l_{jR} -(Y_L)_{ij} \psi_{iL}^T C i \tau_2 \delta^+ \psi_{jL}
+ {\rm h.c.}\biggr]\,,
\label{Lminimal}
\end{eqnarray} 
with $\psi_{iL}= (\nu_{iL}~l_{iL})^T$ and  
$H=(H^0~H^-)^T$.

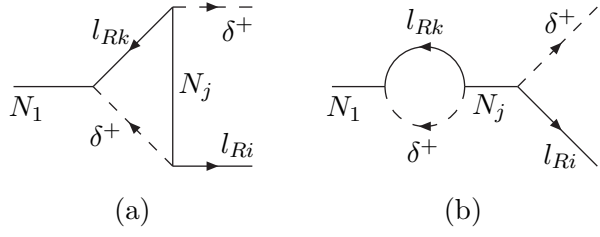
\begin{figure}[t]
\begin{center}
\begin{picture}(200,80)(0,0)
\Line(0,50)(30,50)
\DashArrowLine(60,80)(90,80){5}
\ArrowLine(60,20)(90,20)
\Line(60,80)(60,20)
\ArrowLine(60,80)(30,50)
\DashArrowLine(60,20)(30,50){5}
\Text(5,42)[]{$N_1$}
\Text(35,32)[]{$\delta^+$}
\Text(37,71)[]{$ l_{Rk}$}
\Text(69,50)[]{$N_j$}
\Text(85,72)[]{$\delta^+$}
\Text(85,28)[]{$ l_{Ri}$}
\Text(45,3)[]{(a)}
\Line(120,50)(140,50)
\DashArrowArcn(155,50)(15,0,-180){5}
\ArrowArc(155,50)(15,0,180)
\Line(170,50)(190,50)
\DashArrowLine(190,50)(220,80){5}
\ArrowLine(190,50)(220,20)
\Text(125,42)[]{$N_1$}
\Text(155,26)[]{$\delta^+$}
\Text(156,75)[]{$ l_{Rk}$}
\Text(180,42)[]{$N_j$}
\Text(207,75)[]{$\delta^+$}
\Text(207,24)[]{$ l_{Ri}$}
\Text(170,3)[]{(b)}
\end{picture}
\end{center}
\caption{One-loop diagrams contributing to the lepton asymmetry
in the $N_1$ decay.}
\label{fig}
\end{figure}

We consider the possibility that the scalar singlet is lighter than 
the RH neutrinos and we neglect, at this stage, the 
effects of the $Y_N$ couplings, which are not relevant to 
achieve our main results. We neglect their effects at this stage.
In this case leptogenesis can 
be induced simply by replacing in the diagrams of the standard leptogenesis model, both in the 
loop and in the 
final state, the left-handed lepton doublet with the RH charged 
lepton, $e_R$, $\mu_R$ or $\tau_R$ and the Higgs doublet with the scalar singlet, as shown in Fig.~1.
For the lightest RH neutrino $N_1$, the CP asymmetry, that is to say the average $\Delta_L$ which is produced each time a $N_1$  decays in the thermal bath of the universe at a temperature of order its mass $M_{N_1}$, is:
\begin{equation}
\varepsilon_{N_1}=\sum_i {\frac{\Gamma (N_1 \rightarrow l_{iR} + \delta^+) - 
\Gamma (N_1 \rightarrow 
\bar{l}_{iR} + \delta^-)}{ \Gamma_{N_1}}} \cdot C_L\,,
\label{epsNdef}
\end{equation}     
with its tree level decay width given by
\begin{equation}
\Gamma_{N_1}= \frac{1}{16 \pi} M_{N_1} \sum_i |(Y_R)_{1i}|^2 \,.
\label{width1}\end{equation}
In Eq.~(\ref{epsNdef}), $C_L$ is the lepton number produced in the 
decay $N_1\rightarrow l_{iR}+\delta^+$. 
Unlike the Higgs doublet in the standard 
leptogenesis case, $\delta^+$ does not have a vanishing lepton number. Once 
produced from the decay of the RH 
neutrinos,  it decays
to 2 left-handed antileptons, via the $Y_L$ couplings, so that it has $L=-2$ 
which gives $C_L=-1$. Calculating the one loop diagrams of Fig.~1 one finds
\begin{equation}
\varepsilon_{N_1}=\frac{1}{8 \pi} C_L \sum_j
\frac{{\cal I}m[(Y_R 
Y_R^\dagger)_{1j}^2]}{\sum_i |(Y_R)_{1i}|^2
}\sqrt{x_j}
\left[1-(1+x_j) \log\left(1+\frac{1}{x_j}\right)+\frac{1}{2}\frac{1}{1-x_j}\right] \,,
\label{epsN}
\end{equation}
where $x_j=M^2_{N_j}/M^2_{N_1}$. For this calculation we 
neglected $(M_\delta/ M_{N_1})^2$ corrections which are small as soon as 
the $\delta^+$ is a few times lighter than $N_1$ as we assume here.
In the limit where we also neglect the $M^2_{N_1}/M^2_{N_{2,3}}$ 
corrections, we get
\begin{equation}
\varepsilon_{N_1}=- C_L \frac{1}{8 \pi} \sum_j
\frac{{\cal I}m[(Y_R 
Y_R^\dagger)_{1j}^2]}{\sum_i |(Y_R)_{1i}|^2
}\frac{M_{N_1}}{M_{N_j}}
 \,.
\label{epsNhierarc}
\end{equation}
Apart for the $C_L$ factor and 
for a combinatoric factor of two in the self-energy contribution, 
this asymmetry is the same as in 
the standard case, replacing the ordinary Yukawa couplings $Y_N$ by the
$Y_R$ scalar singlet ones. 
Contrary to the standard case, however, the RH Yukawa 
couplings $Y_R$ do not induce 
any neutrino masses and so are not constrained by them. As a result
this mechanism may easily lead to successful leptogenesis and may also 
work at a much lower scale,
as explained below, which is phenomenologically interesting.

Considering for simplicity only 2 RH neutrinos $N_{1,2}$ (the 
effect of $N_3$ can be straightforwardly incorporated), numerically to 
have successful leptogenesis there are essentially 3 constraints:
\begin{itemize}

\item The total baryon asymmetry produced is given by:
\begin{equation}
\frac{n_B}{s}=-\frac{28}{79} \frac{n_L}{s}=
-\frac{135 \zeta(3)}{4 \pi^4 g_\star}\frac{28}{79}\varepsilon_{N_1} \eta =
-1.36 \cdot 10^{-3}\varepsilon_{N_1} \eta \,,
\label{asymconstr}
\end{equation}
The factor $-28/79$ is the lepton to baryon number conversion factor due to the 
effects of the sphalerons of the Standard Model. $n_B$, and $n_L$ and $s$ 
are the baryon number, lepton number and entropy densities. The 
factor $g_\star=108.75$ appears  to take into account the fact that 
only a fraction $\sim1/g_\star$ of the entropy is due to the 
decaying $N_1$. $\eta$ is the efficiency factor to take into 
account the thermalization effects. $\eta=1$ if all $N_1$ decay 
perfectly out of equilibrium and if there are no fast L violating 
scatterings occuring after the $N_1$ decays. $\eta < 1$ if the $N_1$ 
decay partly in thermal equilibrium with the thermal bath. For a maximal 
efficiency, $\eta=1$, the requirement
to reproduce the data (i.e.~$\frac{n_B}{s}=9 \cdot 10^{-11}$) implies
that 
\begin{equation}
Y_R^{(2)}\equiv \sqrt{\left| \frac{{\cal I}m \left[\sum_i(Y_R)_{1i}(Y_R^*)_{2i}\right]^2}
{\sum_i (Y_R)_{1i}(Y_R^*)_{1i}}\right| } \ge 1.3 \cdot 10^{-3} 
\sqrt{\frac{M_{N_2}}{M_{N_1}}}~,
\label{up2}\end{equation}
which means that
at least one of the $(Y_{R})_{2i}$ coupling needs to be 
of order $10^{-3}\cdot\sqrt{ M_{N_{2}}/M_{N_1}}$ or larger.

\item To avoid a suppression of the efficiency associated to the inverse decay of a $l_R$ 
and a $\delta^+$ into a $N_1$, the decay width has to be smaller than the expansion rate of the universe: 
\begin{equation}
\Gamma_{N_1}
< H(T)|_{T=M_{N_1}}= \sqrt{\frac{4 \pi^3 g_\ast}{45}} 
\frac{T^2}{M_{Planck}}\Big|_{T=M_{N_1}}\,.
\label{ooecN}
\end{equation}
Using Eq.~(\ref{width1}), the corresponding upper 
bound on the $(Y_R)_{1i}$ couplings reads
\begin{equation}
{Y}_R^{(1)} \equiv \sqrt{\sum_i |(Y_R)_{1i}|^2} < 3 \cdot 10^{-4}
\sqrt{\frac{M_{N_1}}{10^9{\rm GeV}}} ~.
\label{up1}\end{equation}
Larger values of ${Y}_R^{(1)}$
lead to suppression of the efficiency which, for successful leptogenesis, 
has to be compensated by 
larger values of the $(Y_{R})_{2i}$ couplings in the 
numerator of the asymmetry $\epsilon_{N_1}$.

\item If Eq.~(\ref{ooecN}) is satisfied, the lepton number washout from $\Delta L=2$ 
scattering mediated by a $N_1$, $l_R \delta^+ \leftrightarrow \bar{l}_R \delta^-$, is 
negligible, because its rate is also smaller than the Hubble rate, see e.g.~\cite{bdp}. 
Taking 
values for $(Y_{R})_{2i}$ consistent with Eq.~(\ref{up2}), the 
washout from the same $\Delta L = 2$  
scattering mediated 
by a $N_{2}$ is generically negligible, except  
possibly for $M_{N_1}$ as low as a few TeV (because for fixed values of the Yukawa couplings, the Hubble rate decreases faster than does the scattering rate when $T \simeq M_{N_1}$ decreases). In fact, this effect depends  
on the interplay 
of $Y_R^{(1)}$, $M_{N_1}$, $M_{N_2}/M_{N_1}$ as well 
as of the $(Y_{R})_{2i}$ couplings.
This interplay can be determined from
the Boltzmann equations. Considering them explicitly, we have checked that
even at scales as low as a few TeV, an efficiency of 
order one can be obtained easily (see \cite{th,bhs,pil2,FHM}).

\end{itemize}

Combining the 3 constraints above, successful leptogenesis can be 
achieved in a large region of parameter space.
The scale at which the lepton asymmetry may be produced depends 
on the hierarchy between the $Y_R$ couplings of $N_2$ and $N_1$. 
This can be quantified by combining Eqs.~(\ref{up2}) and (\ref{up1}):
\begin{equation}
\frac{Y_R^{(1)}}{Y_R^{(2)}} < 0.2 \cdot  \sqrt{\frac{M_{N_1}}{M_{N_2}}\frac{M_{N_1}}
{10^9{\rm GeV}}} ~.
\end{equation}
This condition is easily satisfied for $M_{N_1} \simeq 10^{9-15}$ GeV.
When, for example, $M_{N_1}/M_{N_2}\sim 0.1$ 
and $M_{N_1}=10^7$ GeV, 
at least one of the $(Y_R)_{2i}$ couplings needs to 
be about two orders of magnitude larger than the $(Y_R)_{1i}$.
At scale as low as 1-10 TeV the hierarchy needed is more 
substantial, of about 4 
orders of magnitude, but this is not  
unrealistic for Yukawa couplings (the hierarchy 
needed is of the order of the one 
in the SM Yukawa couplings). An example of a set of parameters leading to 
an efficiency of order one
and to a baryon asymmetry in agreement with the observed one is: 
$M_{N_1}=2$ TeV, $M_{N_2}=6$ TeV, 
$(Y_{R})^{max}_{2i}\simeq 4 \cdot 10^{-3}$, $Y_R^{(1)}\simeq 10^{-7}$ 
and $M_{\delta}\simeq 750$~GeV. We find that successful leptogenesis can 
be generated with $M_{N_1}$ as low as $\simeq 1$~TeV and 
with $M_{N_2}$ as low as $\simeq 4$~TeV.\footnote{If there is an 
additional resonance 
effect, $M_{N_2}$ ($\simeq M_{N_1}$) can be lowered 
down to $\sim 1$~TeV as well.}

So far we have neglected the effects of the ordinary $Y_N$ Yukawa couplings which are necessary to induce the neutrino masses.
Switching them on leads to more tree-level and one-loop diagrams, see more details 
in Ref.~\cite{FHM}. At high scales, $M_{N_1} \gtrsim 4 \cdot 10^8$~GeV,  these diagrams can induce successful leptogenesis
just as in the ordinary scenario. At lower scales they can't because the neutrino constraints require
too small values of $Y_{N}$ couplings\cite{bcst,th,di}. But, still in this case, they can have a suppression effect
on the asymmetry produced by the $Y_R$ couplings through the $Y_N$ contribution 
to the tree level decay width in the denominator
\begin{equation}
\Gamma_{N_1}= \frac{1}{16 \pi} M_{N_1} \sum_i |(Y_R)_{1i}|^2 
+ \frac{1}{8 \pi} M_{N_1} \sum_i |(Y_N)_{1i}|^2
\label{GammaN1tot}\,.
\end{equation}
entering in the denominator of Eq.~(\ref{epsN}).
Just as in the standard leptogenesis mechanism, there will be no inverse 
decay washout effect if $N_1$ contributes to light neutrino masses by less 
than $10^{-3}$ eV, that is to say if the solar and atmospheric mass splittings are
dominated by the contributions of $N_2$ and $N_3$. In fact,  
Eq.~(\ref{ooecN}) now implies the constraint (\ref{up1}) as well as
\begin{equation}
\frac{v^2 \sum_i |(Y_N)_{1i}|^2}{M_{N_1}} < 10^{-3} {\rm eV}~.
\label{upN1}\end{equation}
In the opposite case, larger $Y_R$ couplings to $N_2$ and/or $N_3$ are 
required for
successful leptogenesis, in order to increase $\epsilon_{N_1}$ thus 
compensating for the washout factor $\eta < 1$.

\section{Leptogenesis with a right-handed scalar triplet}

As explained above, the $\delta^+$ must be a singlet 
of $SU(2)_L$ in order that the RH neutrinos can decay into it.  It is important to note that 
this doesn't necessarily mean that the $\delta^+$  must also be a singlet of any other gauge group.  
If the theory of particle interactions beyond the Standard Model contains left-right symmetry \cite{lr}, based on  the group $SU(2)_L\times
SU(2)_R\times U(1)_{B-L}$, one simple possibility is that the $\delta^+$ is the charge-one component 
of an $SU(2)_R$ triplet $\Delta_R$. In this case the leptogenesis mechanism 
discussed in section 2 is slightly modified. 
The relevant interactions are:\footnote{Here for simplicity of notation we take the $\delta^0$ as vevless, that 
is, its contribution to RH neutrino masses is already 
reabsorbed in $M_{N_i}$.}
\begin{eqnarray}
{\cal L} &\owns& 
-M^2_\Delta Tr\Delta_R^\dagger \Delta_R 
+ \left[ -\frac{1}{2} M_{N_{i}} N^{T}_{Ri} C N_{Ri} 
- H^\dagger \bar{N}_{Ri}  (Y_N)_{ij} \psi_{jL} \right. \nonumber \\
& & - (Y_\Delta)_{ij} \psi_{iR}^T C i \tau_2 
\Delta_R \psi_{jR} + {\rm h.c.} \biggr]\,,
\label{lagr}
\end{eqnarray} 
with $\psi_{iL}= (\nu_{iL}$~$l_{iL})^T$, 
$\psi_{iR}= (N_{i}$~$l_{iR})^T$, $H=(H^0~H^-)^T$ and 
\begin{equation}
\Delta_R=
\begin{pmatrix}
\frac{1}{\sqrt{2}}\delta^+ & \delta^{++}  \\
\delta^0 & - \frac{1}{\sqrt{2}} \delta^+ 
\end{pmatrix} \,.
\end{equation}
The diagrams in Fig.~1, in this case, can also lead to successful leptogenesis. They lead to the same asymmetry as in Eq.~(\ref{epsN}), and same constraints, 
replacing everywhere the $(Y_R)_{ij}$ couplings by $\sqrt{2}(Y_\Delta)_{ij}$.
In addition, as there is no coupling of the $\Delta_R$ to two 
left-handed leptons,
the $\delta^+$ does not have $L=-2$ as above and $C_L$ is modified.
Since we assume that the $\delta^+$ is lighter than the RH 
neutrinos, the $\delta^+$ cannot decay to two particles but instead to three,
it decays predominantly to a Higgs doublet and 
a lepton-antilepton pair so that the $\delta^+$ has $L= 0$ with an intermediate RH neutrino, which 
gives $C_L=+1$.

\section{Incorporating Right-handed Leptogenesis in Unified Gauge Theories}

\underline{The case of a charged singlet $\delta^+$}: 
in the presence of left-right symmetry, the leptogenesis mechanism above can be successful 
with a $SU(2)_L\times SU(2)_R$ singlet $\delta^+$ since it can couple in an antisymmetric way to 2 
doublets of
RH leptons 
$\psi_{R}=(N~l_R)^T$.
If the minimal left-right group is further extended to a Pati-Salam model,
$\delta^+$ is  accommodated into a $(1,1,10)$-multiplet under $SU(2)_L \times
SU(2)_R \times SU(4)_c$, which couples bilinearly to RH 
fermions $\sim
(1,2,\overline{4})$. In these cases the presence of a $\delta^+$ is not 
better motivated than in the standard model case (i.e.~it is not related 
to the breaking of the Pati-Salam or left-right symmetry, or contributing 
to fermion masses). The Pati-Salam group may be naturally embedded in unified
models based on $SO(10)$, with all fermions in a $16$-dimensional spinor
representation. In this case $\delta^+$ is part of a $120$ Higgs 
multiplet, which
has renormalizable Yukawa couplings to fermions, contributing 
to fermion masses, see e.g.~\cite{fermionmasses}. 

Alternatively, one can consider the $SU(5)$ option for gauge coupling unification.
In this case, leptons are assigned as follows to $SU(5)$ representations: $\psi_L
\in \overline{5}_f$, $l_R^c \in 10_f$ and $N^c \sim 1_f$. In order to introduce
$\delta^+$, one needs to add to the model a 10-dimensional Higgs multiplet, which
has the proper couplings required in section 2 to achieve RH leptogenesis:  $Y_R 1_f
\overline{10}_f 10_H$ and $Y_L \overline{5}_f  \overline{5}_f 10_H$. More details can be found in
Ref.~\cite{FHM}.

\noindent
\underline{The case of a right-handed triplet $\Delta_R$}:  A RH triplet is naturally 
present in left-right models
since the VEV of its neutral component $\delta^0$ provides the correct symmetry
breaking to the Standard Model and, moreover, it gives a Majorana mass to the 
RH neutrinos. In fact, $\Delta_R \sim (1,3,2)$ couples symmetrically to two RH
lepton doublets $\psi_R \sim (1,2,-1)$. In Pati-Salam models, $\Delta_R$ is
contained in the $(1,3,10)$ multiplet which, in turn, belongs to $\overline{126}$
Higgs representation in $SO(10)$.

The minimal left-right model turns out to be able to satisfy all constraints which are necessary
to lead to successful leptogenesis as in section 3 above, except an important one: it is well known that in order to get a non vanishing CP-asymmetry one must have 2 different sources of flavor breaking,
one in the Yukawa couplings and a different one in the right-handed neutrino mass matrix.
However in the minimal left-right model both matrices are proportional to each other since the
RH neutrino masses are induced from the VEV of the $\delta^0$ through the $Y_\Delta$ couplings.
As a result 
the CP asymmetry is simply vanishing. Therefore, for this leptogenesis mechanism to be 
effective we 
need to extend the minimal model in order to distinguish $M_R$ 
from $Y_\Delta$. For example, one may introduce 
a second RH triplet (a second $\overline{126}$ in $SO(10)$),
or consider extra (e.g.~non-renormalizable) sources of RH neutrino mass.
Alternatively, one could resort to the singlet leptogenesis 
mechanism, adding a $(1,1,2)$ Higgs multiplet ($120$ in $SO(10)$ context).

   
\section{Phenomenology of a TeV Scale $SU(2)_L$ Singlet Charged Scalar}

The observation of a light $SU(2)_L$ singlet $\delta^+$ at 
colliders (produced from a photon, e.g.~from Drell-Yan processes) would 
imply that, in the presence of RH neutrinos, the $Y_R$ 
interactions occur naturally.
This would render our leptogenesis 
mechanism as plausible as the standard one. 
Notice that, as explained above, $\delta^+$ would decay predominantly 
into either a charged 
lepton and a neutrino (or eventually to two different Higgs bosons\cite{FHM}).
Moreover the fact that this 
model can work at scales 
as low as the TeV scale opens the possibility to produce 
directly a 
RH neutrino through the relatively  large $Y_R$ couplings 
of the $N_2$ and/or $N_3$, which can have a mass as low as 
few TeV.\footnote{The production of TeV scale RH neutrinos through 
the Yukawa couplings to left-handed leptons has been discussed e.g. in \cite{pp} for LHC, \cite{AG} for a high energy $e^+e^-$ linear collider and
\cite{piga} for an $e\gamma$ collider.} 
This would leave in general no 
other choice for leptogenesis (and baryogenesis) than to be produced at low 
scale below $M_{N_{2,3}}$, as allowed by our 
model.\footnote{A possible exception is the 
case where the observed $N_i$ has suppressed 
couplings to a given flavor, so that it cannot 
washout any preexisting lepton asymmetry associated to that flavor.}

Note also that the $\delta^+$ singlet 
can induce, through its $Y_L$ couplings,
a $\mu \rightarrow e \gamma$ transition with branching ratio
Br$(\mu \rightarrow e \gamma)\approx
(\alpha/48\pi)|(Y_L)_{e \tau} (Y_L)_{\mu \tau}|^2/$ $(M_\delta^4 G_F^2)$ (see
e.g.~\cite{his}). With $M_\delta$ below TeV, a branching ratio of the order of
the experimental limit (Br$(\mu \rightarrow e \gamma)< 1.2 \cdot 10^{-11}$ 
at $90\%$ C.L. \cite{meg}) can 
be easily obtained. Similarly the $Y_R$ couplings can induce this transition
with Br$(\mu \rightarrow e \gamma)\approx
(\alpha/192\pi)|(Y_R)_{i e} (Y_R)_{i\mu}|^2/$ $(M_{N_i}^4 G_F^2)$, where we assumed that the exchange of the RH neutrino $N_i$ gives the main contribution and we neglected $M_\delta/M_{N_i}$ corrections.
In this case the sets of parameters which lead to successful leptogenesis give rise  
to a smaller branching ratio, below $\sim 10^{-17}$, therefore unobservable.

The case of the triplet $(\delta^0, \delta^+, \delta^{++})$ has a 
similar phenomenology for what concerns the production of the $\delta^+$
and $N_{2,3}$. However, here $\delta^+$ does not have 2-body decays. 
In this scenario a $\delta^{++}$ could also be produced 
electromagnetically 
in colliders. As there is no $Y_L$ couplings, the $\mu \rightarrow e \gamma$
process in this case can be induced only through the $Y_\Delta$ couplings,
with 
suppressed branching ratios as for the singlet case with $Y_R$ couplings.


\section{Summary}

We have considered a new mechanism to induce leptogenesis 
successfully, by the decay of the RH neutrino $N_1$ 
to a RH charged lepton 
and a scalar $SU(2)_L$ singlet $\delta^+$.
In the presence of left-right symmetry
the $\delta^+$ may or may not be
a member of an $SU(2)_R$ triplet.
In both versions one achieves successful leptogenesis easily in a similar way.
This mechanism can work at high scale just as ordinary leptogenesis and it can also work at scales 
as low as few TeV with no need of resonant enhancement of the asymmetry. 
Such a low scale realization  
requires that we do make 3 assumptions: RH neutrinos have to be assumed with masses of order
$\sim TeV$, a lighter charged scalar $\delta^+$ has to exist and RH neutrinosYukawa couplings to RH charged leptons must have a hierarchical structure.\footnote{This can be compared with the resonant ordinary leptogenesis at low scale which to be testable must also be based on 3 assumptions \cite{pil2}: light RH neutrinos must be light in the same way, a hierarchical structure of Yukawa couplings has also to be assumed with in addition a special structure (with large (testable) Yukawa couplings which have to cancel each other to give small enough neutrino masses), and a large degeneracy of RH neutrino masses has to be assumed.} 

In grand-unified theories this mechanism
can be realized,  for the singlet case, 
both in SO(10), if there exists a 120 scalar multiplet, and in 
SU(5) with a 10 scalar multiplet. 
The $SU(2)_R$ scalar triplet case can be 
incorporated in SO(10) models with a $\overline{126}$ scalar multiplet. 
However, in this case, in order for leptogenesis to work, the model should 
contain a source of RH neutrino 
masses independent from this $\overline{126}$ representation.

Phenomenologically, the observation of a light $SU(2)_L$ singlet $\delta^+$ at 
colliders would be a strong evidence in favor of our proposal.  The additional 
production of a RH neutrino at few TeV scale,
through the large couplings to RH charged leptons,  
would make the case for low scale leptogenesis.


\end{document}